\documentclass[twocolumn,amssymb,amsmath,
tightenlines,showpacs,floatfix,
superscriptaddress,aps,prb]{revtex4-1}
\usepackage{amsfonts}
\usepackage{bm}%
\usepackage[colorlinks=true,linkcolor=blue,filecolor=blue,menucolor=yellow,urlcolor=blue,citecolor=blue,anchorcolor=blue]{hyperref}
\usepackage{graphicx}
\usepackage{dcolumn}

\begin{document}

\title{Proximity Effects in Conical-Ferromagnet/Superconductor bilayers} 

\author{Chien-Te Wu}
\email{wu@physics.umn.edu}
\author{Oriol T. Valls}
\email{otvalls@umn.edu}
\altaffiliation{Also at Minnesota Supercomputer Institute, University of Minnesota,
Minneapolis, Minnesota 55455}
\affiliation{School of Physics and Astronomy, University of Minnesota, 
Minneapolis, Minnesota 55455}

\author{Klaus Halterman }
\email{klaus.halterman@navy.mil}
\affiliation{Michelson Lab, Physics
Division, Naval Air Warfare Center, China Lake, California 93555}

\date{\today}

\begin{abstract}
We present a study of various aspects of proximity effects 
in $F/S$ (Ferromagnet/Superconductor) bilayers, where $F$
has a  
spiral magnetic texture such 
as that found in Holmium, Erbium  and other materials, and 
$S$ is a conventional 
$s$-wave superconductor. 
We numerically solve the Bogoliubov-de Gennes (BdG) 
equations self-consistently
and use the solutions to compute physical quantities relevant
to the proximity effects in these bilayers. We obtain the relation
between the superconducting transition temperature $T_c$  and
the thicknesses $d_F$ of the magnetic layer by solving the 
linearized BdG equations. 
We find that the $T_c(d_F)$ curves include multiple oscillations. 
Moreover, the system may be reentrant not only 
with $d_F$, as is the case when the magnet is
uniform, but also with temperature $T$:
the superconductivity disappears in certain ranges of $d_F$ or $T$.
The $T$ reentrance reported here occurs when $d_F$
is larger than the spatial period of the conical exchange field. 
We compute the condensation free energies and entropies from
the full BdG equations and find the results are 
in agreement with $T_c$ values obtained
by linearization. The inhomogeneous nature
of the magnet makes it possible for all odd triplet pairing components
to  be induced.
We have investigated their properties and
found that, as  compared to the singlet amplitude, both the 
$m=0$ and $m=\pm 1$ triplet components exhibit long range penetration. 
For nanoscale bilayers, the proximity lengths for 
both layers are also obtained. 
These  lengths oscillate with  $d_F$ and they are found to be
long range on both  sides. These results are shown to be
consistent with recent experiments.
We also calculate 
the reverse proximity effect described by the three dimensional 
local magnetization, and the local DOS, which reveals 
important energy resolved signatures 
associated with the 
proximity effects. 

\end{abstract}


\maketitle

\section{Introduction}\label{introduction}

The emerging field of spintronics has stimulated interest 
in fabricating solid state devices 
that make use of the intrinsic spins as a 
degree of freedom.\cite{cite:zutic} 
Strides have  been made recently 
towards exploiting the spin  variable 
in hybrid ferromagnet/superconductor $(F/S)$ systems. 
Such systems have shown promise for a number
of practical applications, including nonvolatile information storage.
The simplest of such potential devices usually  involve 
layered 
$F/S$ heterostructures. Owing to these 
potentially important spintronic applications, 
research on the fundamental physics of these
systems has received great attention 
in the last decade.\cite{cite:berg,cite:buzdin4} 
The most important basic physics elucidated
by these studies is probably that of
the superconducting proximity effects in $F/S$ nanostructures,\cite{cite:buzdin4} 
which describes the interplay between ferromagnetic and
superconducting order parameters.
Though these two order parameters are rarely found to coexist 
naturally in bulk materials, 
such coexistence can be and has been achieved near the interfaces of  
artificially created  $F/S$ composites. 
Thus,  the subject has become
important not only for its technological applications
but also because of the underlying fundamental physics. 

In elementary treatments,
ferromagnetism is often deemed strictly incompatible 
with $s$-wave superconductivity 
due to their mutually exclusive order parameters. In ferromagnets, 
the exchange field 
tends to cause 
the electronic spins to align 
in the same direction, while
in singlet $s$-wave superconductors, the Cooper pairs are composed of both 
spin-up and spin-down electrons.
These two order parameters seem to naturally oppose  each other. 
In fact, 
in $F/S$ heterostructures 
this competition 
leads to a strong modification of  
the behavior of the superconducting Cooper pair amplitudes. 
When a Cooper pair in $S$ encounters an $F/S$ interface 
and  enters the $F$ region, the momenta of
spin-up  and spin-down electrons that make up the Cooper pair are changed, 
because of
the exchange field in the $F$ region. This
leads to a nonzero center of mass momentum of 
the Cooper pair\cite{cite:demler,cite:khov1} and 
an overall damped oscillating Cooper pair amplitude
in the $F$  side. 
It is because of these two competing order parameters that 
the
oscillations
decay over a relatively short length 
scale which decreases as the exchange field increases. 

These oscillations of the superconducting wavefunctions 
are one of the most salient features governing   
proximity effects in $F/S$ systems and form the basis for switching 
applications that require the manipulation of the 
superconducting transition temperature $T_c$ 
through variation of the experimental parameters.
Due to the
oscillatory nature of the Cooper pair amplitudes,
the dependence of 
$T_c$ on the thickness of the
ferromagnetic layer,  $d_F$, in $F/S$ bilayers 
is oscillatory too. Furthermore,  the interference between the transmitted
pairing wave function through the $F/S$ interface and the reflected one 
from the boundary can become fully destructive: the superconductivity disappears
for a certain $d_F$ range. 
This superconducting reentrant
behavior  with $d_F$  has been found experimentally in Nb/Cu$_{1-x}$Ni$_x$ and
Fe/V/Fe trilayers\cite{cite:garifullin,cite:zdravkov1,cite:zdravkov2} and it
is well understood theoretically.\cite{cite:buzdin4,cite:buzdin2,cite:khusainov,cite:fominov,
cite:buzdin3,hv1,hv2} 

Another important fact about $F/S$ proximity effects 
is the generation of induced triplet pairing correlations.
These can be generated by the presence of spin active
interfaces,\cite{cite:eschrig,eschrig2,hv3} or (and this is the case we
will focus on here) in systems with clean interfaces and 
inhomogeneous $F$ structure\cite{berg86,berg85,lof,hv07,hv08}.
The simplest such cases are $F_1/S/F_2$ or $F_1/F_2/S$ layers in which the magnetizations 
of the two F layers are misaligned.
For  $s$-wave  superconductors, where the orbital 
part of the pair wave function is even, the Pauli principle requires 
the spin part 
to be odd and this would appear to forbid the existence of triplet correlations.
However, triplet correlations that are odd\cite{berez} in frequency 
(or equivalently in time\cite{hv07}) 
can be induced in $F/S$ systems, with $S$ being $s$-wave pairing, without 
violating the Pauli exclusion principle. 

The importance of odd triplet correlations lies in their long range
nature in the magnet, i.e., their proximity lengths can be  
in principle comparable to those found  
in the usual superconducting proximity effects involving 
nonmagnetic metals. 
Since 
the exchange fields tend to align the electronic spins of
the Cooper pair electrons, the proximity length for
singlet pairing  is very short (and
dependent on the magnitude of exchange field).
However, the triplet pairing correlations can involve  electron pairs with
both spins aligned along the local magnetization direction, 
and thus 
be much less sensitive to the mechanism of exchange fields, penetrating 
much deeper in $F$ than their singlet counterparts. 
The possible appearance of both $m=0$ and $m=\pm1$ ($m$ denoting 
the usual spin quantum number) components of  induced 
triplet correlations is controlled by the symmetry of the system 
and by conservation laws\cite{hv08}. 
In  multilayer $F/S$ 
systems, when the $F$ layers 
are magnetically homogeneous
(all exchange fields are along the same direction, the
quantization axis) 
only the total spin projection 
corresponding to the $m=0$ component can be induced.
On the contrary, all three components ($m=0$ and $m=\pm1$) can arise
if the direction of exchange fields differs in the ferromagnets, 
e.g. the exchange fields of $F_1$ and $F_2$ are not aligned in\cite{hv08} $F_1/S/F_2$
or $F_1/F_2/S$ types of trilayers\cite{cvk1}. 
These long-range characteristics of triplet correlations have 
been experimentally
detected in ferromagnetic multilayers by taking advantage of their magnetic
inhomogeneity.\cite{khaire,gu,sprungmann} 

Besides  ferromagnet misalignment, another possibility 
to generate long-range triplet correlations is to use
a ferromagnet with an intrinsic 
inhomogeneous magnetic texture.\cite{volk} 
Such structures are inherent to either 
known elements or chemical compounds.
Examples of this kind of ferromagnets include 
most prominently Ho,\cite{sosnin} which has a spiral magnetic
structure at low temperatures.  A similar spiral magnetic structure 
is found in metallic Erbium\cite{cable},
MnSi thin films\cite{karhu}, and Fe(Se,Te) compounds\cite{shi}.
Indeed it has been experimentally confirmed that long-range
triplet correlations are induced in Nb/Ho/Co multilayers\cite{robinson}
with the periodicity of Ho playing an important role in triplet
supercurrents. 
Superconducting phase-periodic conductance oscillations have also been 
observed in Al/Ho bilayers\cite{sosnin} where the thickness
of Ho is much larger than the penetration length of singlet amplitudes.
This finding can be explained 
in the framework of the triplet proximity effects.
Theoretically, 
the spin-polarized Josephson current in S/Ho/S junctions
has been studied\cite{dost1} via quasi-classical Green function 
techniques. 
The triplet supercurrent in Ho/Co/Ho trilayers was 
also investigated in the diffusive\cite{dost2} and clean\cite{hal2}
regimes.
The long-range effects can however be limited by 
interface quality and impurities.\cite{hal} 
These earlier works show that ferromagnets with an 
intrinsic spiral magnetic 
structure are of particular interest when studying 
superconducting
proximity effects in $F/S$ nanostructures.

It was also recently predicted\cite{cite:cvk} that 
superconductivity 
in conical-ferromagnet/superconductor bilayers can be reentrant  {\it with temperature}, 
in addition to the standard reentrance
with $d_F$ mentioned above.
It was shown via numerical solution of the Bogoliubov-de Gennes
(BdG) equations 
that in certain cases 
superconductivity can exist in a range $T_{c1}<T<T_{c2}$,
where $T_{c1}$ is nonzero.  
This reentrance with temperature is quite different from that found
in ternary rare earth compounds such as ErRh$_4$B$_4$ and HoMo$_6$S$_8$
\cite{cite:fertig,cite:moncton,cite:ott,cite:crabtree,cite:lynn}
where the disappearance of superconductivity below $T_{c1}$ 
results from the onset of long-range ferromagnetism.
In the  bilayers considered in Ref.~\onlinecite{cite:cvk} 
the high $T$ phase and the low $T$  phase are the same. 
The physical reasons that account for the reentrance there 
are attributed to the proximity effects associated with the
interference of Cooper pair amplitudes and the generation of triplet pairing
correlations, resulting in nontrivial competition 
between the entropies and condensation energies.

In this paper, we present results 
for various properties of the proximity effects 
in $F/S$ bilayers, 
where  the $F$ layer has a helical magnetic structure.
We numerically find the self-consistent
solutions to the 
BdG equations\cite{cite:degennes} and use them
to compute important physical quantities. 
By linearizing the BdG equations we 
calculate the critical temperature as
a function of magnet thickness, exchange field strength 
and periodicity, 
and other parameters. 
We then discuss the
effects of varying  the superconductor thickness to 
coherence length ratio. 
We show that depending on the width of
the superconductor, and for a broad range of magnetic strengths,
reentrant behavior as a function of magnet thickness can arise.
We find 
that under certain conditions, the superconductivity can also be
reentrant with temperature, and for larger $d_F$ values than 
previously reported.\cite{cite:cvk}
To clarify these reentrant phenomena, we
investigate the
thermodynamic functions associated with the various ways reentrance can arise.
We find that all components of the odd triplet correlations can be induced 
and we discuss their long range nature.
We  then characterize the important triplet long range
behavior by introducing the corresponding proximity lengths. 
We find that these 
lengths oscillate as a function of $d_F$, and 
depend on details of the magnetic texture.
Reverse proximity effects are also
studied to determine the magnetic influence on the
superconductor: we calculate the local magnetization vector 
revealing greater penetration into S
for weaker
exchange fields.
Lastly, the spectroscopic information is presented
by means of the local density of states,
to demonstrate consistency 
with the $T_c$ results.

\begin{figure}
\includegraphics[width=0.5\textwidth] {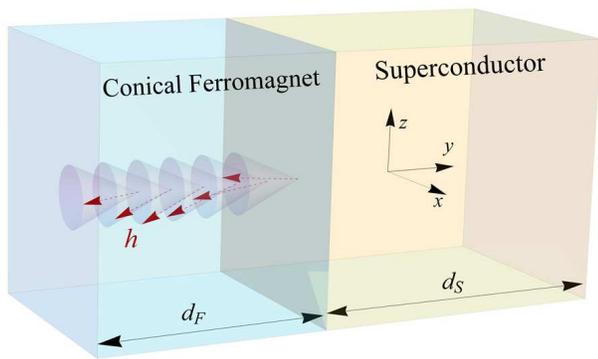}
\caption{(Color online) Diagram of the conical ferromagnet-superconductor
bilayer studied. The spiral 
magnetic structure is described by an exchange field $\bm h$ 
(see Eq.(\ref{exchange})). The 
system is infinite in the $x - z$ plane and finite in $y$. The relevant widths are labeled. }
\label{fig1}
\end{figure}

\section{Methods}

The procedures we employ to self-consistently
solve the BdG equations and to extract the relevant 
quantities are very similar to those already described
in the literature (see Refs.~\onlinecite{cite:cvk,hv3} and references 
therein). It is unnecessary to repeat details here.
We consider $F/S$ bilayers that consist of one ferromagnetic
layer with spiral exchange fields
and a superconducting layer with $s$-wave pairing.
The geometry is depicted in Fig.~\ref{fig1}.
Our systems are infinite in the $x-z$ plane and finite along $y$-axis
Their total thickness  is denoted by $d$:  the $F$ layer has width $d_F$
and the $S$ layer has width $d_S=d-d_F$.
The left end of the bilayers is the $y=0$ plane. 
We assume that the interface lies in the $x-z$ plane 
and the exchange field ${\bf h}$, which is present only in $F$, 
has a component that rotates in this  plane plus 
a constant component in the $y$ direction,
perpendicular to the interface:
\begin{equation}
\label{exchange}
\mathbf{h}=h_0\left\{\cos\alpha\mathbf{\hat y}+\sin\alpha\left[ \sin\left(\frac
{\beta y}{a}\right)\mathbf{\hat x}+\cos\left(\frac{\beta y}{a}\right)\mathbf
{\hat z}\right]\right\},
\end{equation}
where the helical magnetic structure
has a 
turning angle $\beta$, and   
opening angle $\alpha$. 
We will take $a$, the lattice constant, as our unit of length
and vary the strength $h_0$. 
The spatial period of the helix is $\lambda=2 \pi a /\beta$. 

We first write down the effective Hamiltonian of our systems,
\begin{eqnarray} 
\mathcal{H}_{eff} &=& \int
d^3r\Bigl\lbrace\displaystyle\sum\limits_{\rho}\psi_{\rho}^{\dagger}(\mathbf{r})\left(-\frac
{\boldsymbol\nabla^2}{2m^*}-E_F\right)\psi_{\rho}(\mathbf{r})
 \nonumber \\
 &&
 +\frac{1}{2}\left[\displaystyle\sum\limits_{\rho,\rho'}(i\sigma_y)_{\rho\rho'}
 \Delta(\mathbf{r})\psi_{\rho}^{\dagger}(\mathbf{r})\psi_{\rho'}^{\dagger}(\mathbf{r})+
 h.c.\right]
  \nonumber \\
  && -\displaystyle\sum\limits_{\rho,\rho'}\psi_{\rho}^{\dagger}(\mathbf{r})(\mathbf{h}
  \cdot\boldsymbol\sigma)\psi_{\rho'}(\mathbf{r})\Bigr\rbrace,
\end{eqnarray}
where $\rho$ and $\rho'$ are spin indices, $\psi_{\rho}(\mathbf{r})$ is 
the field operator,
and ${ \bm \sigma}$ are the Pauli matrices. 
$\Delta(\mathbf{r})$ in the above equation is the pair potential.
To apply the BdG formalism to spatially inhomogeneous systems, we first 
invoke the generalized Bogoliubov transformation,
\begin{equation}  
\label{transform}
\psi_{\rho}(\mathbf{r})=\sum\limits_n\left[u_{n\rho}(\mathbf{r})\gamma_n+
v_{n\rho}^\ast(\mathbf{r})\gamma_n^{\dagger}\right],
\end{equation}
where $u_{n\rho}(\mathbf{r})$ and $v_{n\rho}(\mathbf{r})$ can be interpreted
as quasi-particle and quasi-hole wavefunctions 
and the creation operator $\gamma_n^{\dagger}$ and annihilation operator $\gamma_n$ 
obey the usual fermionic anti-commutation relations.
To recast the effective Hamiltonian into a diagonalized form, 
$\mathcal{H}_{eff}=\sum_n\epsilon_n\gamma_n^{\dagger}\gamma_n$, one can use 
commutation relations between $\mathcal{H}_{eff}$ and field operators.
By doing so, and making use of the quasi one dimensional
nature of the problem one arrives  at the BdG equations,
\begin{widetext}
\begin{eqnarray}
\begin{pmatrix}
{\cal H}_0 -h_z&-h_x+ih_y&0&\Delta(y) \\
-h_x-ih_y&{\cal H}_0 +h_z&-\Delta(y)&0 \\
0&-\Delta(y)^{\ast}&-({\cal H}_0 +h_z)&h_x+ih_y \\
\Delta(y)^{\ast}&0&h_x-ih_y&-({\cal H}_0-h_z) \\
\end{pmatrix} 
\begin{pmatrix}
u_{n\uparrow}(y)\\u_{n\downarrow}(y)\\v_{n\uparrow}(y)\\v_{n\downarrow}(y)
\end{pmatrix} 
=\epsilon_n
\begin{pmatrix}
u_{n\uparrow}(y)\\u_{n\downarrow}(y)\\v_{n\uparrow}(y)\\v_{n\downarrow}(y)
\end{pmatrix}\label{bogo2},
\end{eqnarray}
\end{widetext}
where ${\cal H}_0\equiv-\frac{1}{2m^*}\frac{\partial^2}{\partial
y^2}+\epsilon_\perp-E_F$, is the usual single particle Hamiltonian
for the quasi one-dimensional problem,
with $\epsilon_\perp$ denoting the kinetic energy associated with the
transverse direction.
By using Eq.~(\ref{transform}), the self consistency 
relation for the pair potential can be rewritten
in the  form,
\begin{equation}
\Delta(y)=\frac{g(y)}{2}{\sum_n}^\prime 
\left[u_{n\uparrow}(y)v_{n\downarrow}^\ast(y)-u_{n\downarrow}(y)v_{n\uparrow}^\ast(y)
\right]\tanh(\frac{\epsilon_n}{2T}),
\label{op}
\end{equation}
where $g(y)$ is the usual BCS superconducting coupling constant in the $S$ region,
and zero in the $F$ material. The prime sign indicates summing over all eigenstates  
with eigenenergies less than or equal to a cutoff ``Debye'' frequency $\omega_D$.
The  solutions to the BdG equations must be determined self-consistently.
This self-consistency condition is extremely important in studying proximity effects.

The singlet pair amplitude, i.e., the amplitude
for  finding a Cooper pair, $F(y)$, is
given by $F(y)\equiv\Delta(y)/g(y)$.
One can determine the superconducting transition temperatures $T_c$ by looking for
the temperatures at which the pair amplitudes becomes vanishingly small.
However, it is much more efficient to find $T_c$ by linearizing  the self
consistency relation, Eq.~(\ref{op}) and using a
perturbation expansion. This technique has been discussed in 
other papers\cite{cite:cvk,pok}
and the details are not repeated here. 

Once  a full set of self-consistent solutions is obtained,
all the additional quantities of interest can be computed.
For example, the triplet correlations,
can in general be defined\cite{hv07,hv08} as:
\begin{subequations}
\label{alltriplet}
\begin{align}
f_0 ({\bf r},t) &\equiv \frac{1}{2}[\langle \psi_\uparrow({\bf r},t)\psi_\downarrow({\bf r},0) \rangle+
\langle \psi_\downarrow({\bf r},t)\psi_\uparrow({\bf r},0) \rangle],
\label{f0def}
\\
f_1 ({\bf r},t) &\equiv \frac{1}{2}[\langle \psi_\uparrow({\bf r},t)\psi_\uparrow({\bf r},0) \rangle-
\langle \psi_\downarrow({\bf r},t)\psi_\downarrow({\bf r},0) \rangle],
\label{f1def}
\end{align}
\end{subequations}
where $\langle \ldots \rangle$ represents ensemble averages.
As discussed in Sec.~\ref{introduction}, both $f_0$ and $f_1$ have to vanish at $t=0$
to comply with the Pauli principle. 
However, in general both of them can be induced when $t\neq0$ and the magnetic
structure is inhomogeneous.
By using Eq.~\ref{transform} and considering the time evolution, one can rewrite the odd
triplet correlations in terms of quasi-particle and quasi-hole wavefunctions
in the form:
\begin{subequations}
\label{alltripleta} 
\begin{align}
f_0 (y,t) & = \frac{1}{2} \sum_n \left[ u_{n\uparrow} (y)
v_{n\downarrow}^{\ast}(y)+
u_{n\downarrow}(y) v_{n\uparrow}^{\ast}(y) \right] \zeta_n(t), 
\label{f0defa} \\
f_1 (y,t) & = \frac{1}{2} \sum_n \left[ u_{n\uparrow} (y)
v_{n\uparrow}^{\ast}(y)-
u_{n\downarrow}(y) v_{n\downarrow}^{\ast}(y) \right] \zeta_n(t),
\label{f1defa}
\end{align}
\end{subequations}
where $\zeta_n(t) \equiv \cos(\epsilon_n t)-i \sin(\epsilon_n t) \tanh(\epsilon_n /(2T))$.

Given the quasi-particle amplitudes and eigenvalues, 
one is also able to evaluate the thermodynamic
quantities from the free energy $F(T)$. For an inhomogeneous
system it is most convenient to use the expression:\cite{cite:kosztin}
\begin{align}
F(T) = -2T \sum_n \ln \left[2 \cosh \left(\frac{\epsilon_n}{2T}\right )
\right]+\left \langle \frac{\Delta^2(y)}{g(y)} \right \rangle_s,
\label {fe}
\end{align}
where here $\langle \ldots \rangle_s$ denotes spatial average. 
The condensation free energy $\Delta F$ is 
the difference between the free energies of the superconducting state,
$F_S$, and the normal states $F_N$, i.e. $\Delta F=F_S-F_N$. 
$F_N$ can be calculated by assuming that the pair potential 
is absent throughout the system.

Another important physical quantity, which
can be determined experimentally 
by  tunneling spectroscopy, is the local density of
states (LDOS). 
This quantity often reveals important information about 
the superconducting features of the
sample studied. In our quasi-one-dimensional model, 
the LDOS $N(y,\epsilon)$  depends spatially only on $y$.
$N(y,\epsilon)$ consists of both spin-up and spin-down LDOS
contributions, that is,
$N(y,\epsilon)=N_{\uparrow}(y,\epsilon)+N_{\downarrow}(y,\epsilon)$.
\begin{align}
\label{dos}
N_\rho(y,\epsilon) = \sum_n [|u_{n\rho}(y)|^2 
\delta(\epsilon-\epsilon_n)+ |v_{n\rho}(y)|^2 \delta(\epsilon+\epsilon_n)], 
\end{align} 
where $\rho = \uparrow, \downarrow$. 

Just as  the superconducting order parameter is 
changed by the presence of ferromagnets,
also, near the interface, the ferromagnetism can be modified by the 
presence of the superconductor,\cite{cite:khov1,fryd,koshina,bergeretve,us69}
a phenomenon known as the reverse proximity effect. 
It is best described by
considering the local magnetization $\mathbf{m}$.
The local magnetization vector is  defined as 
${\bf m}=-\mu_B\langle\sum_{\sigma} \Psi^\dagger {\bm \sigma} \Psi \rangle$, 
where $\Psi \equiv (\psi_\uparrow, \psi_\downarrow)^T$, 
and, again, it depends on the coordinate $y$ only.
By using Eq.~\ref{transform}
we have,
\begin{subequations}
\label{mm}
\begin{align} 
m_x(y) =& - \mu_B \sum_{n}\biggl\lbrace (u_{n\uparrow}^{\ast}(y)u_{n\downarrow}(y)
        +u_{n\downarrow}^{\ast}(y)u_{n\uparrow}(y)) f_n\nonumber \\
        & +(v_{n\uparrow}(y) v_{n\downarrow}^{\ast}(y)+v_{n\downarrow}(y) v_{n\uparrow}^{\ast}(y)) (1- f_n)\biggr\rbrace,\\
m_y(y) =& i \mu_B \sum_{n}
\biggl\lbrace (u_{n\uparrow }^{\ast}(y)
u_{n\downarrow}(y)
        -u_{n\downarrow}^{\ast}(y)u_{n\uparrow}(y)) f_n\nonumber \\
        & +(v_{n\uparrow}(y) v_{n\downarrow }^{\ast}(y)-v_{n\downarrow}(y) 
	v_{n\uparrow}^{\ast}(y)) (1- f_n)\biggr\rbrace,\\
m_z(y) =& -\mu_B \sum_{n}\biggl\lbrace (|u_{n\uparrow}(y)|^2- |u_{n\downarrow}(y)|^2) f_n \nonumber \\
&+(|v_{n\uparrow}(y)|^2- |v_{n\downarrow}(y)|^2) (1- f_n)\biggr\rbrace, 
\end{align}\end{subequations}
where $f_n$ is the Fermi function of $\epsilon_n$ and $\mu_B$ is the Bohr magneton.

\section{Results}

In the results shown 
here, 
capital letters will always denote the
dimensionless lengths denoted by the
corresponding small letter. 
For example, the dimensionless thickness of the ferromagnet
is written as $D_F \equiv d_F/a$ and that of superconductors
is $D_S \equiv d_S/a$, where $a$ is the lattice constant 
in Eq.~(\ref{exchange}). 
For the helical magnetic structure we take angular 
values  (see Eq.~(\ref{exchange}))
$\alpha=4\pi/9$ and $\beta=\pi/6$ which 
are\cite{sosnin,cite:linder} appropriate 
to  Ho, 
in which case 
$D_F=12$ 
contains one full period of the spiral exchange field.
We will denote this dimensionless
spatial period by $\Lambda$ in the following subsections.
For materials other than Ho many of the results can be read
off by rescaling $\Lambda$ to the appropriate value. 
Throughout this paper, the dimensionless
superconducting coherence length is fixed to be $\Xi_0=100$.
In the same spirit, the dimensionless exchange field, $I$, is
measured in terms of the Fermi energy: $I\equiv h_0/E_F$. We 
choose the  Fermi wavevector in $S$ to equal $1/a$. We take
the ``Debye'' cutoff value to be $\omega_D=0.04E_F$. As usual, 
this value is irrelevant except for setting the overall
transition temperature. Temperatures are given in
dimensionless form in terms of $T_c^0$, the transition
temperature of bulk  $S$ material.
When discussing the triplet amplitudes, which are time dependent,
we use the dimensionless  time $\tau\equiv t \omega_D$.
Vertical dashed lines shown in figures, when
present, denote the $F/S$ interface. 

\begin{figure}
\includegraphics[width=0.5\textwidth] {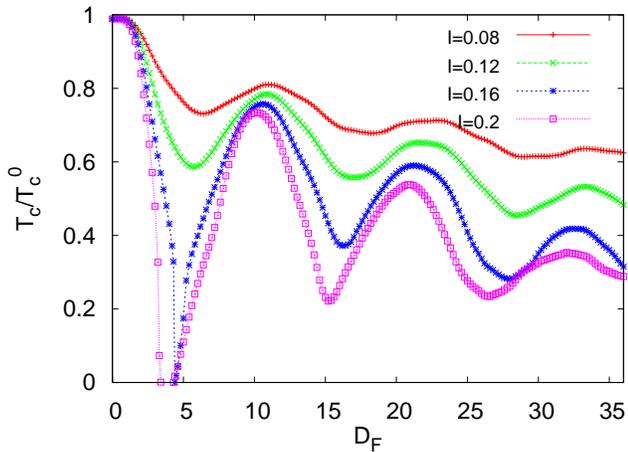}
\caption{(Color online) Calculated transition temperatures $T_c$, normalized to
$T_c^0$, vs. $D_F$ for several values of 
the dimensionless  exchange field $I$ (see text). In this figure $D_S$ is fixed
for all values of $I$ to be $1.5\Xi_0$. The lines 
connecting data points are guides to the eye.}
\label{fig2}
\end{figure}

\begin{figure}
\includegraphics[width=0.5\textwidth] {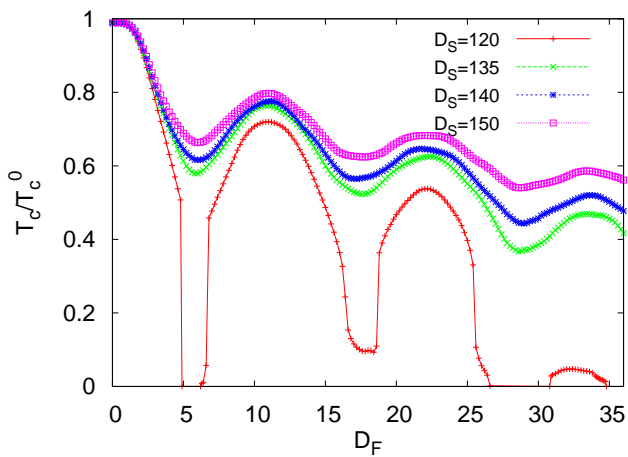}
\caption{(Color online) Transition temperatures $T_c$ 
vs. $D_F$ at $I=0.1$
and several $D_S$.  The lines are
guides to the eye.}
\label{fig3}
\end{figure}

\subsection{Transition Temperatures}

To investigate  the details of the predicted 
oscillatory nature of the $d_F$ dependence 
of $T_c$ as discussed in Sec.~\ref{introduction}, we 
calculated $T_c$ as a function of $D_F$ for several $I$ and 
$D_S$ values. These results are shown in Fig.~\ref{fig2} and Fig.~\ref{fig3}.
The $D_F$ range in both figures
includes three complete periods of the spiral magnetic order.
This is reflected in the results shown:
indeed, the presence of multiple  oscillations
in the included range of $D_F$
is the  most prominent feature in Figs.~\ref{fig2} and \ref{fig3}.
The oscillations in $T_c$ arise (as we discuss
below) from a combination
of the periodicity of the spiral 
 magnetic structure and the usual
$T_c$ oscillations which arise, even when the magnet is 
uniform, from the difference\cite{cite:buzdin4,cite:demler} in the
wavevectors of the up and down spins. 
In Fig.~\ref{fig2}, one can also see that with stronger exchange fields 
the oscillation amplitudes are larger.
Despite this increase of the amplitudes with the
exchange field (they are approximately proportional to
$I$), the overall $T_c$ decreases when
the exchange field increases.
This is consistent with  expectations: a stronger exchange field
destroys the superconductivity more efficiently.
As mentioned in Sec.~\ref{introduction}, 
when the exchange field is strong enough, the systems can become
normal in some range $D_{F1}<D_F<D_{F2}$. 
Indeed, reentrance with $d_F$  can be seen to occur in Fig.~\ref{fig2} near 
$D_F=4$ at $I=0.2$. 
Another feature seen in this figure is the decrease
of the amplitude oscillations with increasing $D_F$.
This arises simply because the singlet Cooper pair amplitudes in $S$ 
near the $F/S$ boundary decay more strongly at larger
$d_F$ and therefore the  effect
of the pair amplitude oscillations in $F$  is weaker.\cite{zhu}  

In a $F/S$ bilayer where the ferromagnet is homogeneous, 
the periodicity of the $T_c$ oscillations
is governed by the exchange field, or equivalently,
by the magnetic coherence length\cite{cite:khov1}  $\Xi_F=1/I$. 
Here, where a bilayer
with a conical inhomogeneous ferromagnet is considered,  
the intrinsic spiral magnetic order with spatial period
$\Lambda$ plays an equally
important and competing  role in the $T_c$ oscillations.
In other words, both the strength and the periodicity
of exchange fields influence the overall decay
and the oscillatory nature of the superconducting transition 
temperatures. 
The existence of two different 
spatial periodicities leads to the
obvious consequence that the
$T_c(D_F)$ curves are not describable in terms
of one single period.
However, when $I$ is not very strong ($I \lesssim 0.1$), the minima of $T_c$ 
are near the locations where $D_F=\Lambda/2$, $3\Lambda/2$, 
and $5\Lambda/2$ and similarly, the  $T_c$ maxima occur near $D_F=\Lambda$, 
$2\Lambda$, and $3\Lambda$. This
indicates that the magnetic
periodicity is dominant. 
Roughly speaking, the maxima and the minima are correlated
with the strongest and weakest spatial average
of the exchange field components  in
$F$. 
As $I$ increases and
$\Xi_F$ decreases deviations become obvious.
Figure~\ref{fig2}  shows that the 
distances between two successive maxima decrease 
when the exchange fields increase. 

The existence  of the multiple oscillations 
discussed above
has been confirmed experimentally.
In Ref.~\onlinecite{chiodi}, 
 $T_c$ in Nb/Ho bilayers was measured as 
a function of $d_F$. The results exhibit an overall
decay with Ho thickness, on which there
are superimposed  oscillations which are correlated
with, but not simply described by, the spatial
wavelength $\lambda$ of the Ho structure. Comparison
with 
the theory discussed here was made, 
using $I$  as an adjustable parameter. Values
near $I=0.1$ were found to provide the best fit.
The other parameters were extracted from other
known properties of Ho and Nb or (e.g. $d_S$) from
the experimental sample geometry.
The results of the comparison were extremely
satisfactory, showing clear agreement
in all the features of the rather intricate $T_c(d_F)$
experimental curves. It was found also that one
of the samples was close to being
reentrant with $d_F$ at a value very close to 
that
predicted by theory.

\begin{figure} 

\includegraphics[width=0.5\textwidth] {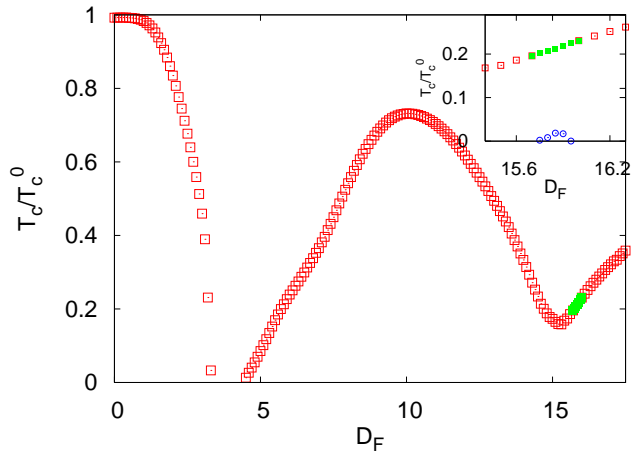} 
\caption{(Color online) Calculated transition temperatures $T_c$ 
vs. $D_F$ for $D_S=148$ and $I=0.2$.   
The main plot
shows ((red) symbols) the overall $T_{c}$ behavior from 
$D_F=0$ to $D_F=1.5\Lambda$.
Reentrance with $D_F$ near $D_F=4$ is seen. In this case there is
also reentrance with temperature in the
region indicated by (green) symbols near $D_F$=16.
The inset is a blow up of this region:  superconductivity
exists only in the region $T_{c1}<T<T_{c2}$, where  $T_{c2}$
is depicted by the upper (green) squares and
$T_{c1}$ by the (blue) circles.  }  
\label{fig4}
\end{figure}

In Fig.~\ref{fig3}, we present $T_c$ results for several 
values of $D_S$
ranging from $1.2\Xi_0$ to $1.5\Xi_0$, with a fixed exchange field
$I=0.1$.
One can  see that the distance between successive maxima
is an extremely weak function of $D_S$.
This agrees with our previous discussion: the oscillatory nature in $T_c$ is
chiefly dependent on the exchange fields and magnet structure.
Since superconductivity is more robust for larger $D_S$, 
the ferromagnet lowers the overall $T_c$ for thinner superconductors
as evidenced in Fig.~\ref{fig3}.
Fig.~\ref{fig3} also demonstrates that not only a strong $I$ can lead
to $D_F$ reentrances, but also a thinner $D_S$.
Interestingly, at the
smallest value of $D_S$ considered, there are two $D_F$ reentrance regions,
one 
near $D_F=5$ and the other near $D_F=27$.
As discussed above, 
these $D_F$ reentrances in both Fig.~\ref{fig1} and ~\ref{fig2}
are mainly due to the interference 
between the transmitted and reflected Cooper pair condensates
that are oscillatory in the F region.
%

In  previous work\cite{cite:cvk}, we reported one
specific case where superconductivity in Ho/S bilayers
exhibits not only the usual reentrance with $d_F$ but also,
at some fixed values of $d_S$, $h_0$, and $d_F$, 
reentrance with 
$T$, that is, superconductivity  
exists only  in a temperature
range $T_{c1}<T<T_{c2}$, where $T_{c1}$ is finite.
In the example reported in Ref.~\onlinecite{cite:cvk}
temperature reentrance occurred   
near the first minimum of the $T_c(D_F)$ curve.
We have investigated here whether 
this kind of reentrance can occur near 
some of the other minimum of $T_c(D_F)$. 
These locations appear favorable for such an occurrence 
since superconductivity is relatively weak near these minima.
Also reentrance with $D_F$ is 
after all an extreme case of a minimum $T_c(D_F)$.
We have found
that other $T$-reentrant examples can indeed be found, 
although by no
means universally.
Here we report an example of reentrance occurring near 
the second minimum of $T_c(D_F)$.
In Fig.~\ref{fig4}, the main plot shows $T_c(D_F)$ 
for the parameter values specified in the caption. The first 
minimum of $T_c(D_F)$
drops to zero and is an example of $D_F$ reentrance. 
In the
region near the second minimum ((green) symbols) reentrance
with $T$  occurs. The region of interest is enlarged in the
inset. There 
the upper (green) symbols represent $T_{c2}$ and
the small dome of lower (blue) circles  represent $T_{c1}$. 
In the dome region, but not outside it,
the superconductivity is reentrant in $T$.
When one lowers the temperature from  the
normal region, the $F/S$ bilayers
become superconducting at $T_{c2}$. With further cooling,
the bilayers return to normal state.  
Reentrance in this case occurs at the second minimum 
rather than the first because there is no upper transition
associated with the first minimum: the system is normal.
Near the second minimum the oscillatory effects 
are not as strong, and as as a result,
the system becomes reentrant in $T$.
This can be viewed as a ``compromise'':
near the second minimum, as opposed to the first, 
superconductivity is not completely                        
destroyed but it becomes ``fragile''
and can disappear upon lowering $T$. The physics involved
from the thermodynamic point of
view will be discussed in the following subsection.

\subsection{Thermodynamics of Reentrance phenomena}

\begin{figure}
\includegraphics[width=0.5\textwidth] {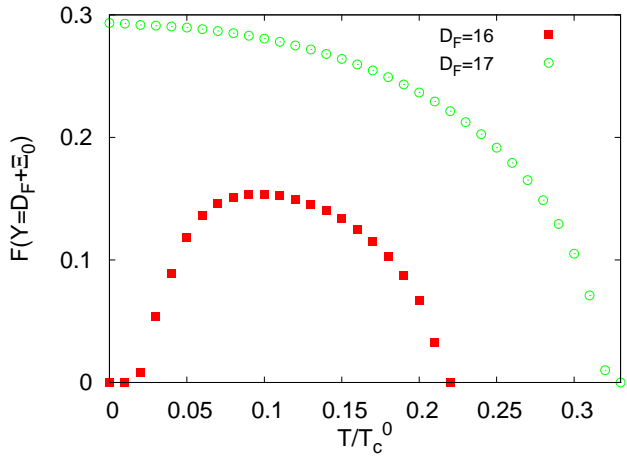} 
\caption{(Color online) The singlet pair amplitude, normalized to its value
for bulk $S$ material, at a location one coherence length inside $S$
from the Ho/S interface, plotted as a function of $T$. The (red)
squares are for
$D_F=16$ and the (green) circles are for $D_F=17$.
All other parameters are as in Fig.~\ref{fig4}.} 
\label{fig5}
\end{figure}

To understand the reentrance phenomena in $T$ it is most useful
to examine the thermodynamics of the two transitions, and in the region
between them. 
From the condensation free energy $\Delta F$, which can be evaluated
as explained in connection with Eq.~(\ref{fe}), other quantities such
as the condensation energy and entropy are easily obtained.
For reentrance with
$D_F$, it is sufficient to look at the free  energy at constant low $T$.

Considering  the reentrance with $T$, 
it is illuminating to consider first the $T$ dependence of the singlet
pair amplitude $F(Y)$ well inside the $S$ material. Thus, we
focus on $F(Y)$  
one coherence length from the  interface: $Y=D_F+\Xi_0$.
This quantity, normalized to its value in bulk $S$ material,
is plotted in Fig.~\ref{fig5} as a function of $T$ 
for two contrasting 
values of $D_F$, one at $D_F=16$ where reentrance occurs (see Fig.~\ref{fig4})
and at a very nearby value, $D_F=17$, which lies just 
outside the reentrance region and exhibits typical behavior. 
Fig.~\ref{fig5}  ((green) circles) demonstrates that 
in the latter case the amplitude
behaves qualitatively as the order parameter does
in a conventional BCS superconductor: it decreases very
slowly near $T=0$ and eventually drops to zero
very quickly but continuously  near $T_c$, 
indicating the occurrence of a second
order  phase transition. This transition occurs at $T_c/T_c^0=0.32$ 
in agreement with Fig.~\ref{fig4}.
However, the behavior of the pair amplitude in the
reentrant region ((red) squares in Fig.~\ref{fig5}) is quite different.
There are two transition temperatures: below  a very low but finite
temperature, $T_{c1}/T_c^0=0.02$, the singlet pair amplitude   
vanishes and 
the system is in its normal state. $F(Y)$ then begins
to rise continuously, has a maximum at a temperature 
$T_m$, (where $T_m/T_c^0 \approx 0.1$)
and eventually drops to zero, again continuously, at an upper 
transition $T_{c2}/T_c^0 \approx 0.22$. In the region $T_{c1}<T<T_{c2}$ 
the system is in the superconducting state.
Both transitions are of second order. The values of
$T_{c1}$ and $T_{c2}$  from the vanishing of the
amplitude, seen in Fig.~\ref{fig5}, 
agree with those calculated
directly from linearization of the self
consistent equation
plotted in Fig.~\ref{fig4}. 

\begin{figure}
\includegraphics[width=0.5\textwidth] {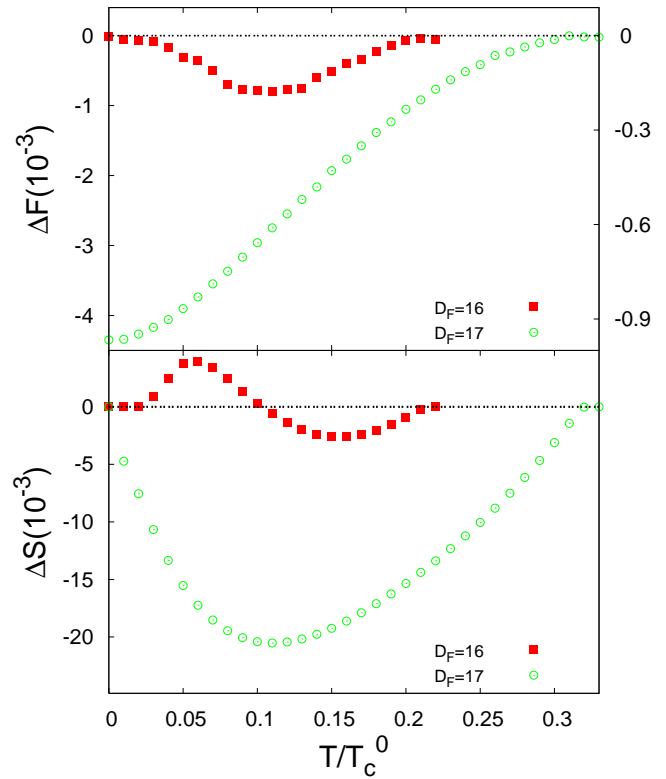} 
\caption{(Color online) The normalized condensation free energies,
$\Delta F=F_S-F_N$, vs. $T/T_c^0$ are shown in the top panel 
for the same cases presented in Fig.~\ref{fig5}. The (red) squares
and right scale are for $D_F=16$. The
(green) circles and left scale are for $D_F=17$. 
The bottom 
panel shows the normalized (see
text) entropy differences, $\Delta S=S_S-S_N$
vs. $T/T_c^0$, on the same vertical
scale. The meaning of the symbols is the same as in the top panel.} 
\label{fig6}
\end{figure}

We now turn to the condensation free energy, $\Delta F$, 
and  entropy, $\Delta S$, for the same $T$ reentrant case. 
$\Delta F$ is shown in the top panel of Fig.~\ref{fig6} 
as calculated from Eq.~(\ref{fe})  
and normalized by $2E_0$, where $E_0$ is the 
condensation energy of bulk $S$ material
at $T=0$. The lower panel shows the
normalized condensation entropy, defined as the
$\Delta S \equiv -d\Delta F/d(T/T_c^0)$. 
The meaning of the symbols in this figure
is the same as in the previous one.
When the system is near (but outside) the reentrant
region the behavior of both quantities plotted is qualitatively
the same as that found in textbooks for bulk BCS superconductors.
Quantitatively, the magnitude of $\Delta F$ for our systems
are much smaller than that for  bulk $S$  where we would have 
$\Delta F=-0.5$ at $T=0$ in our units. 
The value of $T_c$ in the non-reentrant case can also be identified
from where 
the free energies of the normal and
superconducting states are the same ($\Delta F(T) \equiv 0$), 
and it agrees with both Fig.~\ref{fig4} and ~\ref{fig5}.
Moreover, the vanishing 
of the entropy difference at a finite
$T_c$ confirms the 
occurrence of a second order phase transition. 
The value of this transition temperature is consistent 
with all above results.

\begin{figure} 
\includegraphics[width=0.5\textwidth] {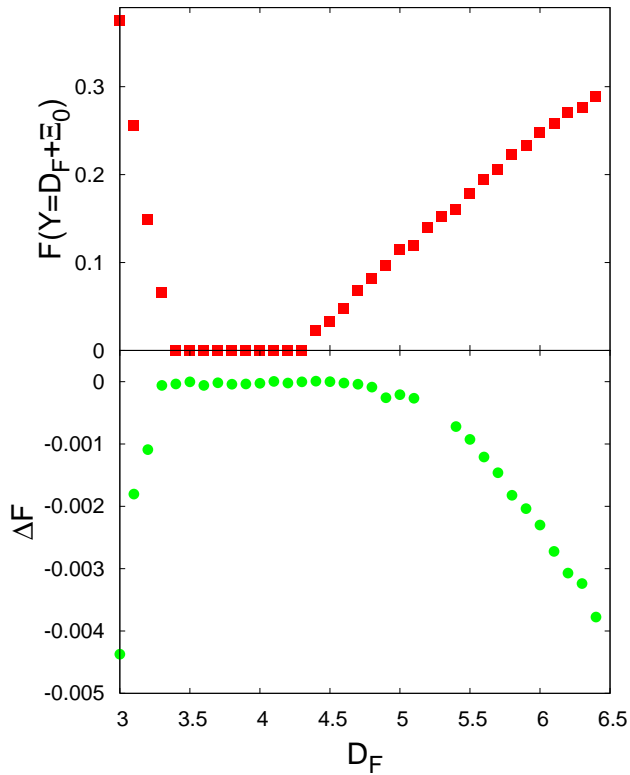}
\caption{(Color online) Reentrance with $D_F$. 
Top panel: normalized singlet pair amplitude, 
computed at a location one coherence length inside $S$
from the  interface,  as a function of $D_F$. Bottom
panel:  normalized condensation free energy, 
$\Delta F=F_S-F_N$, vs. $D_F$  at $T=0$.}
\label{figdf}
\end{figure}

The story for the reentrant case
is quite different. There, although the values of  $\Delta F$
are much smaller
compared to those in
the standard case, one can still find that the
minimum of $\Delta F$ occurs at approximately the same value 
$T_m$ where the singlet pair amplitudes have a maximum. Thus
the superconductivity is most robust at $T=T_m$. 
The two transition temperatures $T_{c1}$ and $T_{c2}$ 
can also be determined from the top panel of Fig.~\ref{fig6} and match with 
those found in Figs.~\ref{fig4}
and ~\ref{fig5}.
In the two $T$ ranges
$T<T_{c1}$ and $T>T_{c2}$, the normal state is the only 
self-consistent solution to the 
basic equations, as is evident from Fig.~\ref{fig5}.  
The vanishing $\Delta F$ when $T<T_{c1}$ means that
the electrons do not then condensate into Cooper pairs.
This is exactly what happens for pure superconductors when $T>T_c$.

There are some remarkable facts about the behavior of $\Delta S$
in the reentrant case.
First, the vanishing of $\Delta S$
(along with that of $\Delta F$) in Fig.~\ref{fig6} indicates
that the system undergoes second order phase transitions at
both $T_{c1}$ and $T_{c2}$.
Also, $\Delta S$ is positive for $T_{c1}< T< T_m$ where
$T_m$ is again the
value of $T$ at which the singlet pair amplitude reaches
its maximum {\it and} $\Delta F$
its minimum. That the entropy of the superconducting state
is higher than that of the normal state indicates that the normal state
at $T_{c1}<T<T_m$ is {\it more} ordered
than the superconducting one.
This truly unusual fact, which is the root
cause of the reentrance, is due to the
oscillating nature of both the Cooper pair condensates and of the
exchange field, which leads to an uncommonly
complicated structure for the pair amplitude.
Above $T_m$, the superconducting state becomes more ordered than
the normal state: $\Delta S$ is negative.
From Fig.~\ref{fig5} and ~\ref{fig6}, we see that the singlet pair
amplitudes, the condensation free energies, and entropy differences
of reentrant case in the range $T_m<T<T_{c2}$
have a similar trend to those of the non-reentrant
case in the range $0<T<T_{c}$. 
We have found also examples of
non-reentrant cases in which there is a finite temperature $T_m$
at which $\Delta F$ has a minimum but where on further lowering $T$, $\Delta F$
remains negative all the way to $T=0$. 

\begin{figure*} 
\includegraphics[width=1\textwidth] {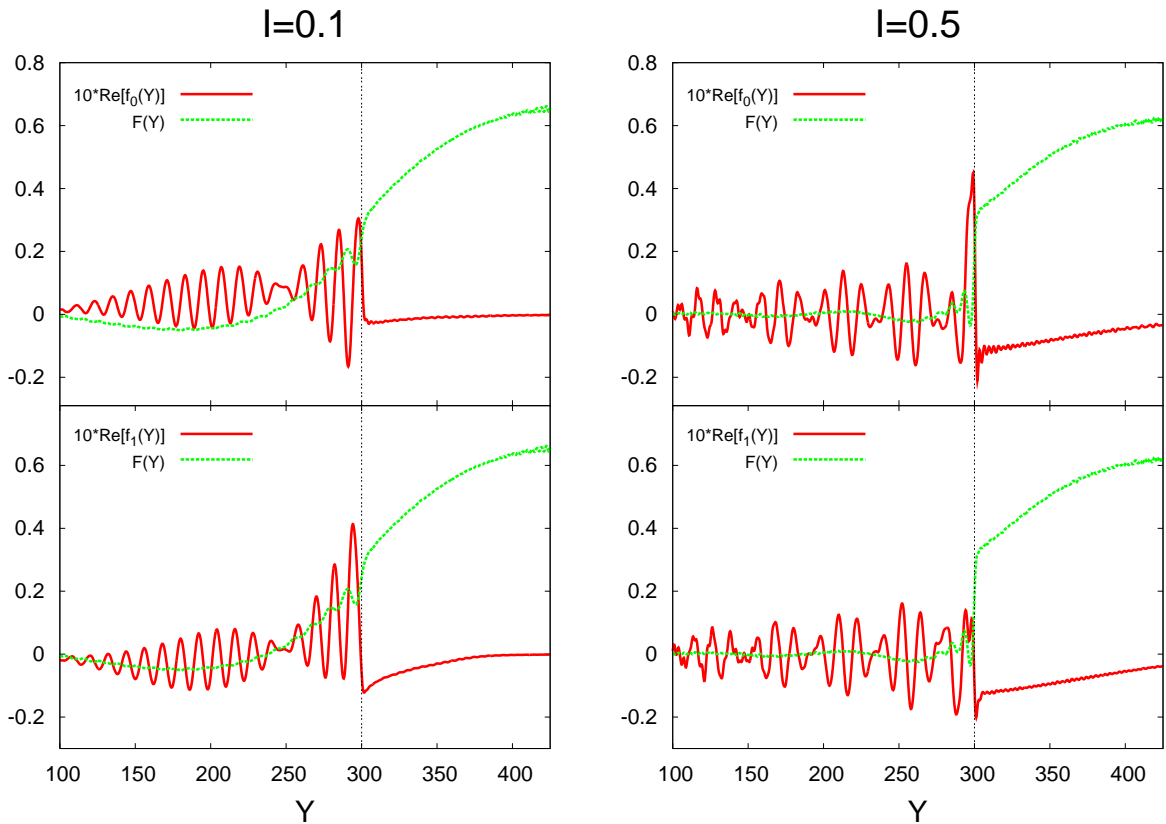} 
\caption{(Color online) Comparison between the
spatial dependencies of the singlet pair amplitude 
and  the induced triplet correlations 
at the two indicated values of $I$,  at $D_S=150$, $D_F=300$
and $T=0$.
The $S$ region is to the right of the dashed vertical line.
Both singlet, $F(Y)$ ((green) curves
higher in the $S$ region))  and triplet, $f_0(Y)$, $f_1(Y)$,
pair amplitudes are
normalized to the  value of $F(Y)$ in pure bulk $S$ material. For this
comparison, the normalized induced triplet pair amplitudes,
which are evaluated  at $\tau=9.6$, 
are multiplied
by a factor of $10$.  The real parts of $f_0(Y)$ and $f_1(Y)$
are shown ((red) curves strongly oscillating in the $F$ region)).
}
\label{fig7}
\end{figure*}

\begin{figure}
\includegraphics[width=0.5\textwidth] {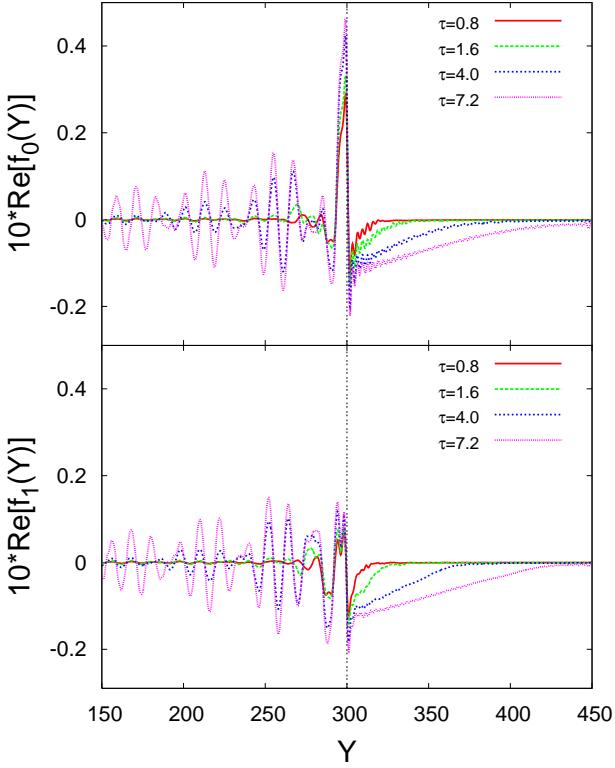} 
\caption{(Color online) Real parts
of the induced triplet pair amplitudes, normalized as
in the previous figure, for different characteristic times $\tau$.
In these plots, $D_S=150$, $D_F=300$
and $I=0.5$. The top panel shows the real part of $f_0(Y)$ and the bottom one
that of $f_1(Y)$.}
\label{fig8}
\end{figure}

\begin{figure*}
\includegraphics[width=1\textwidth] {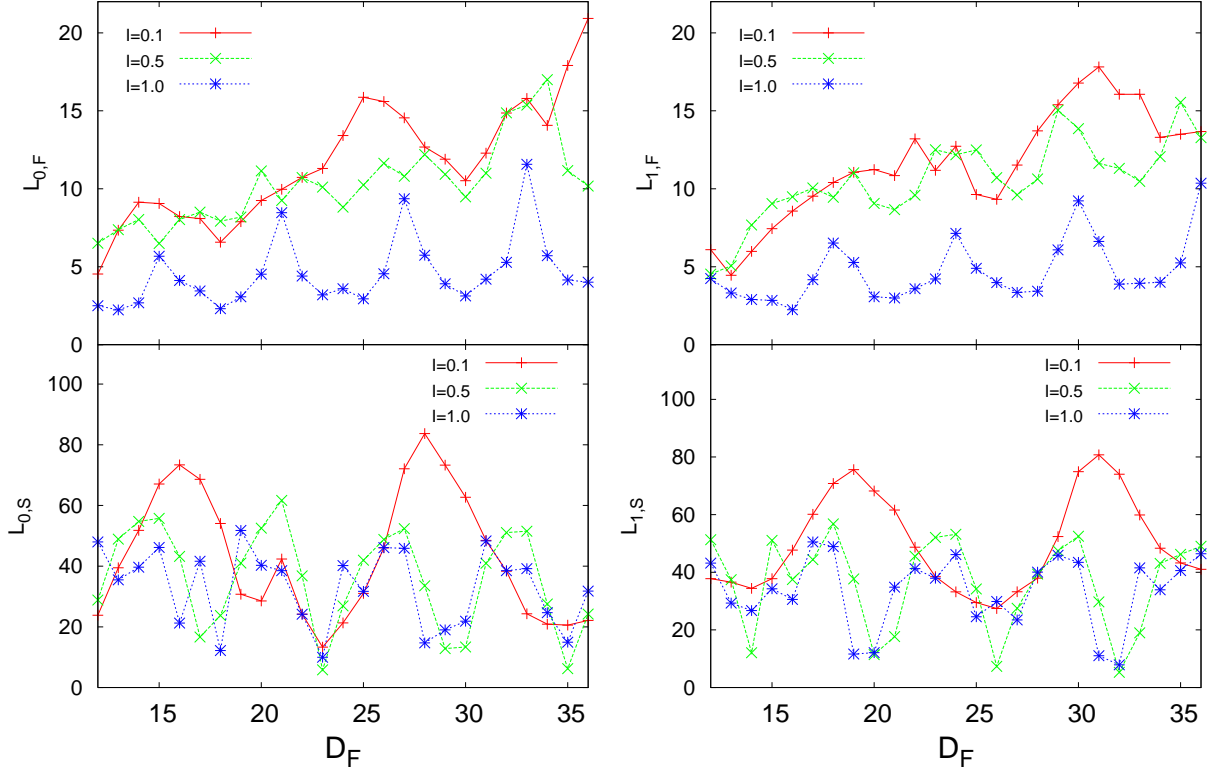}
\caption{(Color online) The proximity lengths $L_{i,M}$
(see Eq.~(\ref{pen})  of the induced triplet pair amplitudes
vs. $D_F$ for different $I$, at $\tau=4.0$ and $D_S=150$. The left 
panels show the proximity lengths $L_{0,F}$ and
$L_{0,S}$ (from $f_0$ in
the $F$ and $S$ regions) and the right panels $L_{1,F}$
and $L_{1,S}$, similarly extracted from $f_1$. 
The lines are guides to the eye.}
\label{fig9}
\end{figure*}

\begin{figure*}
\includegraphics[width=1\textwidth] {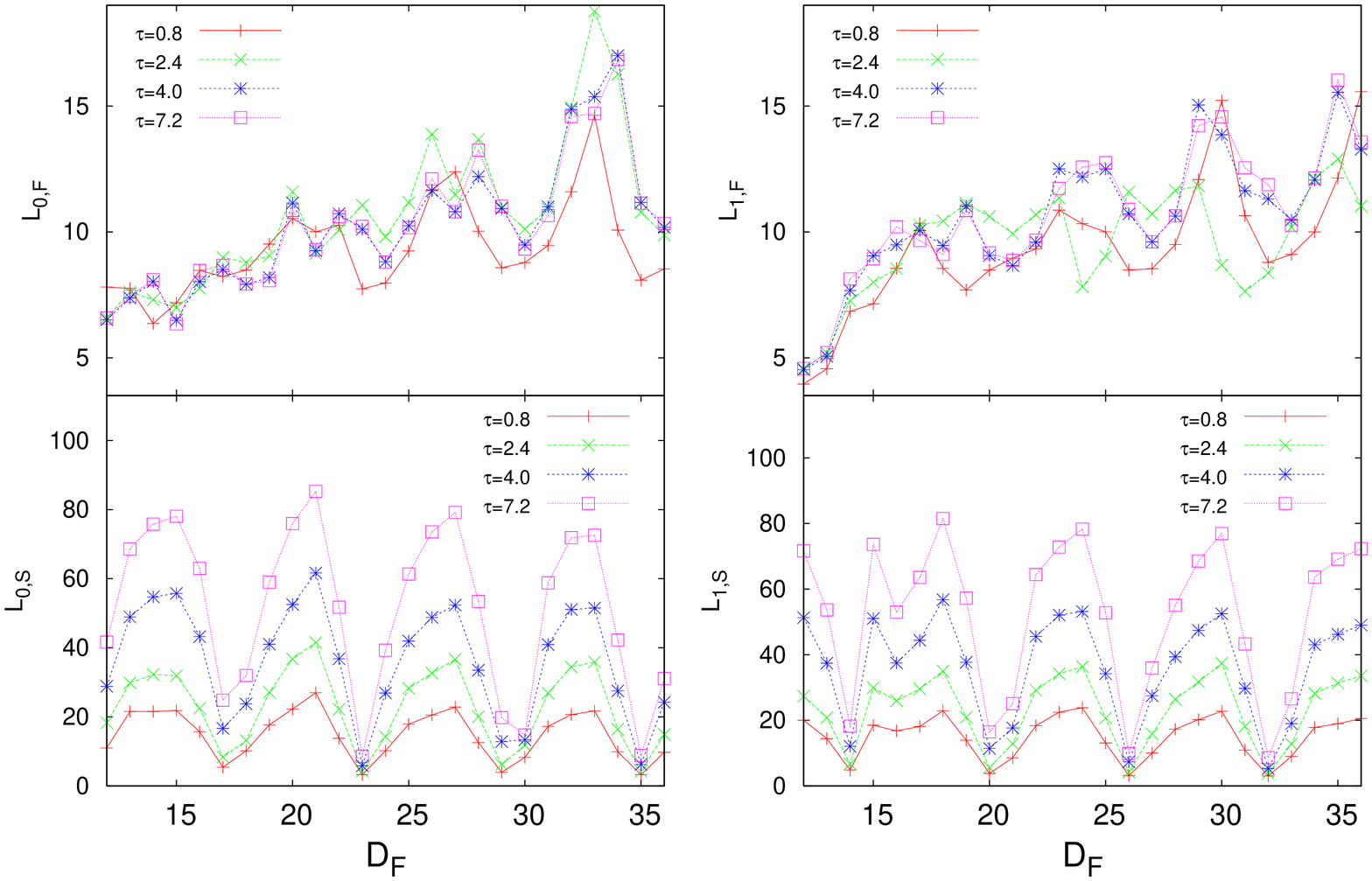}
\caption{(Color online) Triplet proximity lengths
vs. $D_F$ for $I=0.5$, $D_S=150$, at different $\tau$ values. 
The left panels show $L_{0,F}$ and $L_{0,S}$,
and the right panels show $L_{1,F}$  and $L_{1,S}$.
The lines are guides to the eye.}
\label{fig10}
\end{figure*}

The situation in the more
common $D_F$ reentrance region, where we find that the system
does not become superconducting
when it is heated from $T=0$, 
is different from that of $T$ reentrance. 
A case where $T_c$ vanishes in the range $D_{F1}<D_F<D_{F2}$ for $I=0.2$ 
was seen in Fig.~\ref{fig2}. 
To further analyze this $D_F$ reentrance, we again calculated 
the singlet pair amplitudes inside $S$ 
at one coherence length from the interface,  
in the zero temperature limit. 
The top panel of Fig.~\ref{figdf} shows the normalized $F(D_F+\Xi_0)$
as a function of $D_F$, for the same parameters as 
the $I=0.2$ case in Fig.~\ref{fig2}.
The singlet amplitudes drop to zero 
in the same range as where $T_c$ vanishes in Fig.~\ref{fig2}:
the normal state is the only self-consistent solution 
and the superconductivity is completely destroyed in this $D_F$ range.
One can also see that the order parameter is continuous but
its derivative  discontinuous at $D_{F1}$ and $D_{F2}$. 
In the bottom panel, we plot the corresponding 
condensation free energies (at $T=0$) as 
a function of $D_F$. The $D_F$ range and the temperature
are the same as the top panel. 
The condensation free energies vanish in the same 
$D_F$ reentrance region. 
Unlike the  derivatives of the singlet pair amplitudes, 
the derivatives of $\Delta F$ at $D_{F1}$ and $D_{F2}$ 
appear to be continuous. 

The physical origins of these two kinds of reentrance are 
not identical.
As mentioned in Sec.~\ref{introduction}, 
the interference effects of oscillating Cooper pair wavefunctions are 
responsible for the $D_F$ reentrance, provided that $I$ is strong
and $D_F$ is not too thick. $D_F$
reentrance does not require a nonuniform magnet.
The conical-ferromagnet structure introduces an additional
nonuniform magnetic order which may
coexist with nonuniform superconductivity 
as predicted in Ref.~\onlinecite{cite:anderson}. 
This additional non-uniformity, with its
concomitant  introduction of triplet correlations
and of a new periodicity, can produce, as we have
shown, reentrant behavior in $T$, as opposed
to the simpler behavior seen e.g. near the first minimum in the main plot 
of Fig.~\ref{fig4}. 
Thermodynamically, the reentrance with $T$ is due to the competition
between entropy and energy,\cite{cite:cvk} and driven by the high entropy
of the disordered superconducting state. 
When $T<T_m$, $\Delta S$ is positive 
and the roles of the normal and superconducting 
phases are exchanged: the high entropy 
phase is the superconducting one. 
Further lowering $T$ brings the system back to normal state.
One can  compare the instance of $T$ reentrance reported here with 
that reported in our 
previous work\cite{cite:cvk}
where it occurs near the first minimum of $T_c(D_F)$. 
In that work, $D_S=150=1.5\Xi_0$ and $I=0.15$. Here, not 
only is $D_S$  thinner but also
$I$ is greater. The first minimum of $T_c(D_F)$ in the main
plot of Fig.~\ref{fig4}
drops to zero and becomes a $D_F$ reentrance region. 
Because stronger $I$ and thinner $D_S$ are unfavorable
to superconductivity, the system can not sustain $T$ reentrance there.
Thus, a delicate balance of geometrical and material parameters is
required.

\subsection{Singlet to triplet conversion}

In this subsection, we will discuss the general properties of 
the induced triplet pairing correlations in $F/S$ bilayers with
$F$ being a conical ferromagnet. 
As mentioned in Sec.~\ref{introduction}, 
in the presence of inhomogeneous exchange fields 
in the $F$ layers 
both the $m=0$ and the $m=\pm 1$ triplet pair amplitudes        
are {\it allowed} by conservation laws and the
Pauli principle, but this says {\it nothing} about their size or shape, or
indeed on whether they will exist at all. Thus detailed calculations 
are needed. 
The intrinsically inhomogeneous 
magnetic textures discussed here provide unique
opportunities to study
the triplet proximity effects in
$F/S$ systems containing only 
a single $F$ layer.
Triplet correlations in the ballistic regime for both $F_1/S/F_2$ and 
$F_1/F_2/S$ trilayers  have been found
in previous work\cite{hv07,hv08,cvk1} to be long ranged
and the expectation\cite{robinson} that they will also
be in our case is fulfilled. 
We will here discuss and characterize
this and other aspects (such as the effect of the 
strength of the exchange fields on the triplet
pair amplitudes)
of triplet pairing correlations in $F/S$ bilayers 
where the magnets maintain
a spiral exchange field. 
Results presented in this subsection are all
in the low $T$ limit. 

To exhibit the long range nature of both types
of triplet amplitudes, 
we show in Fig.~\ref{fig7} both the triplet and singlet pair
amplitudes for a thick $F$ layer as a function of position,
as given by the dimensionless coordinate $Y$.
In  this (and the next figure, Fig.~\ref{fig8}),
we will focus  on the real parts of the in general
complex (see Eqs.~(\ref{alltripleta})) $f_0$ and $f_1$, 
since we have found
that, for the cases shown, their imaginary parts are
smaller by at least a factor of 2 to 5 and 
their behavior is similar to that of the real parts. 
To properly 
compare singlet and triplet quantities,
both the singlet amplitude, $F(Y)$ and the
triplet amplitudes are normalized the same way: to the
value of the singlet amplitude in bulk  $S$ material.
For visibility, we have multiplied 
the triplet pair amplitudes 
by a factor of $10$.
In the left and right columns
of Fig.~\ref{fig7},  we show in this way the real parts of both $f_0$ and $f_1$ 
when $I=0.1$ and $I=0.5$, respectively. 
The ferromagnet  has a large thickness: $D_F=3\Xi_0=2D_S$.
The triplet correlations, which we recall vanish
at equal times, are computed at 
a value of the dimensionless  time  $\tau=9.6$. 
One sees right away that both the $f_0$ and $f_1$ components
can be induced at the same time. This is always
the case in our structures, as opposed
to what  occurs
in $F_1/F_2/S$ and $F_1/S/F_2$ trilayers where 
the $f_0$ and $f_1$ components can be induced simultaneously only when
the exchange fields in these $F$ layers are non-collinear.
Secondly, the induced triplet correlations on the
$F$ side are long ranged compared
to the singlet amplitudes. The singlet amplitudes decay 
with a short\cite{cite:khov1} 
proximity length $2\pi\Xi_F \approx 2\pi/I$ due to the 
pair-breaking effect of the exchange field.
In contrast, the proximity length for the 
triplet amplitudes as seen in  Fig.~\ref{fig7} 
is much longer: it is of the order of $\Xi_0$, and does not
depend strongly on $I$. The triplet amplitudes  
spread over the $F$ side with an oscillatory behavior.
This difference is more pronounced in the $I=0.5$ case, where the decay 
length $\Xi_F$ is much shorter than  $\Xi_0$ and the singlet amplitudes
diminishes much faster than in the  $I=0.1$ case. 
For both $I=0.1$ and $I=0.5$, one can also see that the 
singlet amplitudes begin to rise from the $F/S$  
interface and saturate in 
the $S$ side about one superconducting coherence length from
the interface.
This agrees with our previous work\cite{cite:khov1}.
Another interesting feature seen in the $I=0.5$ case is that the peak height of 
the $f_1$ component near the interface is not much higher than 
that of its other peaks, as happens with
its $f_0$ counterpart.
In other words, the subsequent peak heights in the $F$ regions 
are comparable to that of the peak nearest to the interface.

In delineating the role of triplet correlations
in other experimentally relevant quantities,
it is necessary to understand their time dependence.
Due to the self consistent
nature of the proximity effects
and the fact that the
triplet condensate amplitudes
are odd in time, their time dependence
is in general nontrivial. 
We illustrate this in Fig.~\ref{fig8}, where we show the spatial
dependence of both the $m=0$ and $m=\pm 1$ components
of the triplet amplitude for several $\tau$. 
The parameters  used here are the same
as in the right panels ($I=0.5$) of Fig.~\ref{fig7}.
For easier comparison with Fig.~\ref{fig7}, we have again multiplied 
the normalized triplet amplitudes by a factor of $10$.
Figure~\ref{fig8} shows that at  small times  triplet correlations
are generated only near the  interface.
(We have of course verified that they always vanish when $\tau=0$).
One can  extract information about the proximity length from the growing
increase of peak heights in the $F$ regions.
The peak heights grow faster when they are deeper 
inside the ferromagnet. 
Moreover, Fig.~\ref{fig8} clearly demonstrates that 
the triplet correlations penetrate into $F$ regions 
as $\tau$ increases, in the range studied.
More remarkably, the peaks of the $f_1$ component that are
not nearest to the interface
grow very fast in time and have heights that are comparable to the one
nearest to the interface, consistent with our remarks 
in our discussion of Fig.~\ref{fig7}.
In contradistinction with the oscillating behavior of the triplet amplitudes
in the $F$ regions,
one can see that both $f_0$ and $f_1$   decay monotonically into
the $S$ side without any oscillations.
However, the triplet correlations still spread over in the $S$ regions
at larger values of $\tau$ just as  they do in the $F$ layer.

In the above paragraphs, we have discussed the long range nature and other
properties of the triplet amplitudes in our system
when the conical ferromagnet is very thick.
In the following paragraphs, we will consider the proximity effect
of induced triplet pairing correlations 
for smaller scale conical-ferromagnets. 
To quantify the effect we introduce
a set of proximity lengths $L_{i,M}$ defined as:
\begin{align}
\label{pen}
L_{i,M}=\dfrac{\int_M dY|f_i(Y,\tau)|}{\max_M|f_i(Y,\tau)|}, \quad i=0,1 \quad
M=S,F.
\end{align}
Here the first index denotes 
the spin component, and
the second index $M$  denotes the 
region in which the given function is evaluated.
If the decays were exponential, these lengths would
coincide with the characteristic length in the exponent. 
Obviously, in the present situation the decays are more 
complicated but the $L_{i,M}$ 
can easily be extracted numerically.
They depend on $D_S$, $D_F$, $I$, and $\tau$. 
The range of $D_F$ we will
consider is from $\Lambda$ to $3\Lambda$. 
In Fig.~\ref{fig9}, we plot these proximity lengths on both the $F$
and $S$ sides for three different values of  $I$, at $\tau=4.0$.
The left panels show the $f_0$ proximity lengths and 
the right panels that extracted from $f_1$ .
Recall that $I=1$ corresponds to the half-metallic limit.
We consider first the $F$ side (top two panels)
One can clearly see that 
both $L_{0,F}$ and $L_{1,F}$ are correlated to 
the strength of the exchange fields. 
Fig.~\ref{fig9} displays a  period 
of near $\Lambda/2$ for  
both $L_{0,F}$ and $L_{1,F}$ 
at $I=1.0$. 
We also see that the peak heights  increase only slowly with
increasing $D_F$.
Also, the locations of the maxima or minima
of $L_{0,F}$  are  locations of minima or maxima, respectively,
of $L_{1,F}$. 
This is as one might expect from the rotating character
of the field.
On the other hand, 
for  $I=0.1$ or $I=0.5$ 
the periodicity is not clear
since, for reasons
already mentioned, the intermingling of 
periodicities becomes more complicated.
Overall the proximity lengths  are larger 
than those in the half-metallic limit.
However, one can still say that both $L_{0,F}$
and $L_{1,F}$
gradually increase, although with fluctuations, with 
$D_F$.

\begin{figure}
\includegraphics[width=0.5\textwidth] {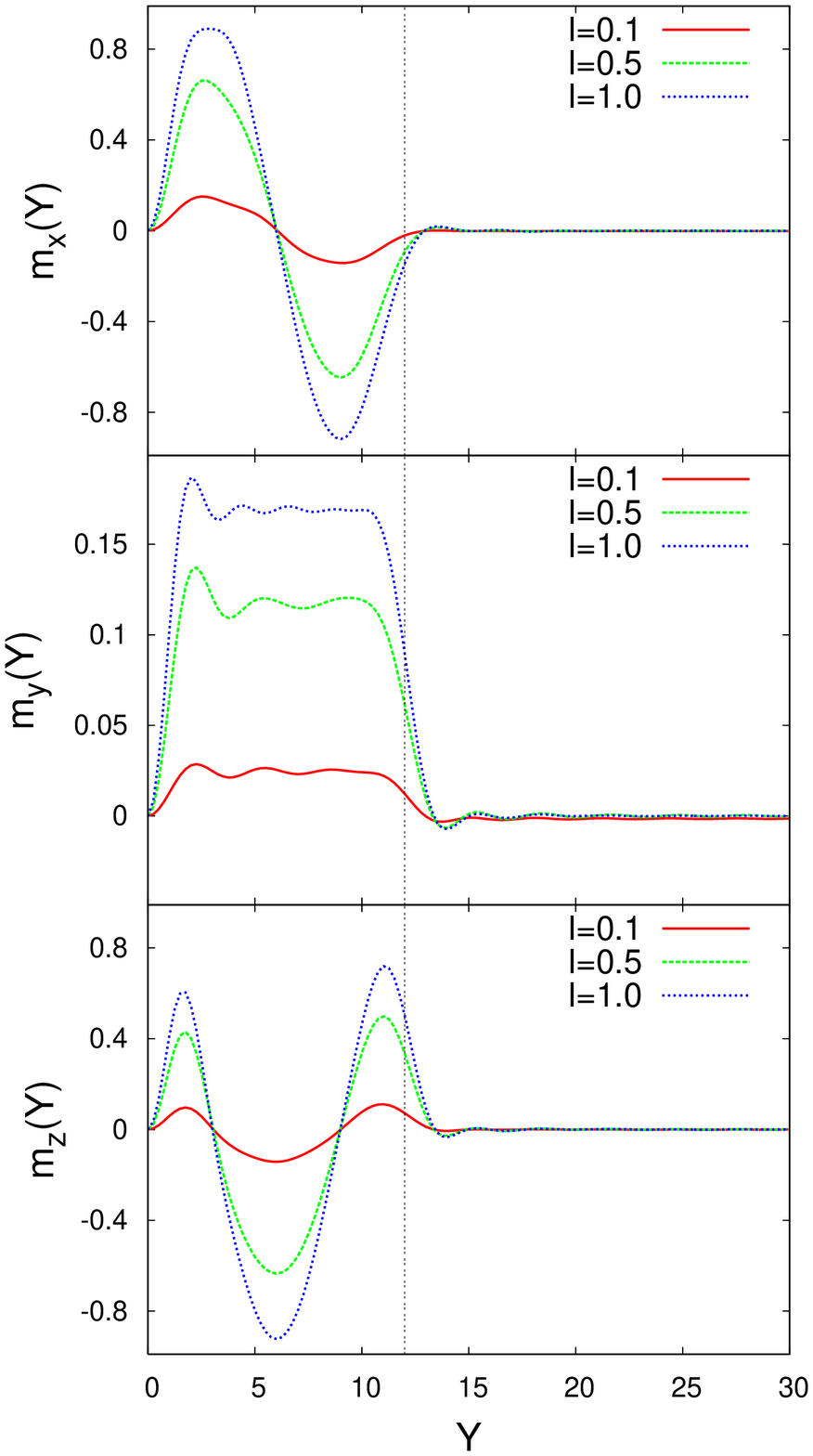}
\caption{(Color online) Normalized (see text)
local magnetization components plotted as
a function of $Y$ for several 
values of $I$. From top to bottom, $x$, $y$, and
$z$-components are shown. We use $D_F=\Lambda$ and $D_S=150$ in this figure.}
\label{fig11}
\end{figure}

The superconductor on the 
$S$ side is intrinsically $s$-wave but 
because of the $F$ layer, triplet
correlations can be induced in it, near the interface,
as seen in Figs.~\ref{fig7} and \ref{fig8}. Their decay, which
is now monotonic, can be equally characterized by
the proximity lengths defined in Eq.~(\ref{pen}). Results are plotted in
the bottom panels of Fig.~\ref{fig9}. 
The minimum of $L_{0,S}$ is, for all
three  values of $I$, 
at $D_F=23$ which is near $2\Lambda$. 
The maxima of $L_{0,S}$ for $I=0.1$ 
are at $D_F=16$ and $D_F=28$ which are
not far from $1.5\Lambda$ and $2.5\Lambda$ respectively.
On the other hand, for  $L_{1,S}$ 
the maxima for $I=0.1$ are at $D_F=19$
and $D_F=31$ and there is a minimum  at $D_F=26$. 
The locations of these maxima are still near $1.5\Lambda$ and
$2.5\Lambda$ and they are only slightly different than what they are
for $L_{0,S}$ case.
If one recalls the above  discussion of Fig.~\ref{fig3},
maxima of $T_c$ occur 
when $D_F$ is close to an integer  multiple of $\Lambda$.
Since a higher $T_c$ is correlated with a higher singlet pair amplitude, 
this suggests again that there exists a 
conversion between singlet and triplet 
Cooper pairs.
The dependence of  $L_{0,S}$ and $L_{1,S}$,  at $I=1.0$, 
on  $D_F$ is harder to characterize. This
is because the high value of
$I$ reduces the scale of the overall proximity effect in $S$ 
(i.e. the depletion
of the singlet amplitude).  
At $I=0.5$, one  still finds
that the approximate periodicity of 
$L_{0,S}$ and $L_{1,S}$ 
is
about $\Lambda/2$. 
The proximity lengths $L_{0,S}$ and $L_{1,S}$ 
are again anti-correlated
at $I=0.5$: the maxima (minima) locations of $L_{0,S}$
are near the minimum (maximum) locations of $L_{1,S}$.

Recent experiments\cite{robinson} 
in systems that consist of two superconducting
Nb electrodes coupled via a Ho/Co/Ho trilayer
have revealed that the long range effect
of triplet supercurrents was much more
prominent at particular thicknesses
of the Ho layers.
The magnetic coherence length in Ho in the experiment
was $\sim 5$ nm which would correspond in our notation 
to $I\sim 0.1$. \cite{chiodi}
In the experiment, the Ho thickness was symmetrically varied
and the critical current, $I_c$, at $T=4.2K$ was measured.
Peaks of $I_c$ corresponding to $D_F=0.5\Lambda$
and $D_F=2.5\Lambda$ were found.
These experimental findings are consistent with our theory.
Here, we have shown (see Fig.~\ref{fig9}) that $L_{1,S}$ has maxima
near $1.5\Lambda$ and $2.5\Lambda$
when $I=0.1$ in the $D_F$ range we have considered.
We found another maximum at $D_F\sim 0.5\Lambda$,
not included in the range shown.
The penetration lengths associated with 
$S$ are as important as 
those associated with $F$ when discussing the triplet proximity effect, 
because the system can open up the corresponding channels 
only when both of them 
are long ranged.
We believe that no obvious peak near $1.5\Lambda$ was observed
because of the layout of their symmetric system. 
Therefore, one can conclude that the spiral magnetic structures
play an important role in the triplet proximity effects.
Both experiment and theory confirm that the existence
of the long range proximity effects depends on the relation between 
the thickness of the magnetic layers and the wavelength of their
magnetic structure.

Having seen in the previous two figures
that  triplet amplitudes may substantially  
pervade even rather thick Ho layers
at moderate values of $\tau$,
it is of interest to 
investigate the $\tau$ dependence of the
proximity lengths in these nanoscale $F/S$ systems 
for times roughly up to $2\pi$, in our dimensionless units. 
We therefore present 
in Fig.~\ref{fig10}, the triplet proximity lengths 
as a function of $D_F$ for $I=0.5$, and 
at different values of $\tau$. 
The panel arrangement is as
in the previous figure. Thus,
in the top panels where we plot $L_{0,F}$ and $L_{1,F}$,
we see that both of them 
depend only weakly  on  $\tau$, in the range considered.
This is in part because of  the 
relatively thin $F$ layers included in the plot.
The triplet amplitudes vanish at $\tau=0$ but
can saturate quickly through the $F$ region as soon
as $\tau$ increases. 
In contrast, on the much thicker ($D_S=150$) 
$S$ side (bottom
panels) both  $L_{0,S}$ and $L_{1,S}$ 
increase with $\tau$, as is
consistent with expectation and
previous work involving $F/S$ 
systems with misaligned exchange fields.\cite{hv08} 
Furthermore, the overall shape of the proximity 
lengths on the $S$ side does not change with $\tau$ and only
the magnitude evolves.
Quite remarkably, the minima of $L_{0,S}$ and $L_{1,S}$ 
are very deep, and the
value of these lengths at their minima is
almost $\tau$ independent and nearly the same
at all minima in the range plotted. The
minima are separated by $\Lambda/2$.
If one compares the left and right panels 
one can see that the locations of maxima in one approximately
coincide with the position of minima in the other:
the left and right
panels are again complementary to each other
as was the case with the plots in Fig.~\ref{fig9}.

\subsection{Local magnetization and LDOS}

\begin{figure*}
\includegraphics[width=1\textwidth] {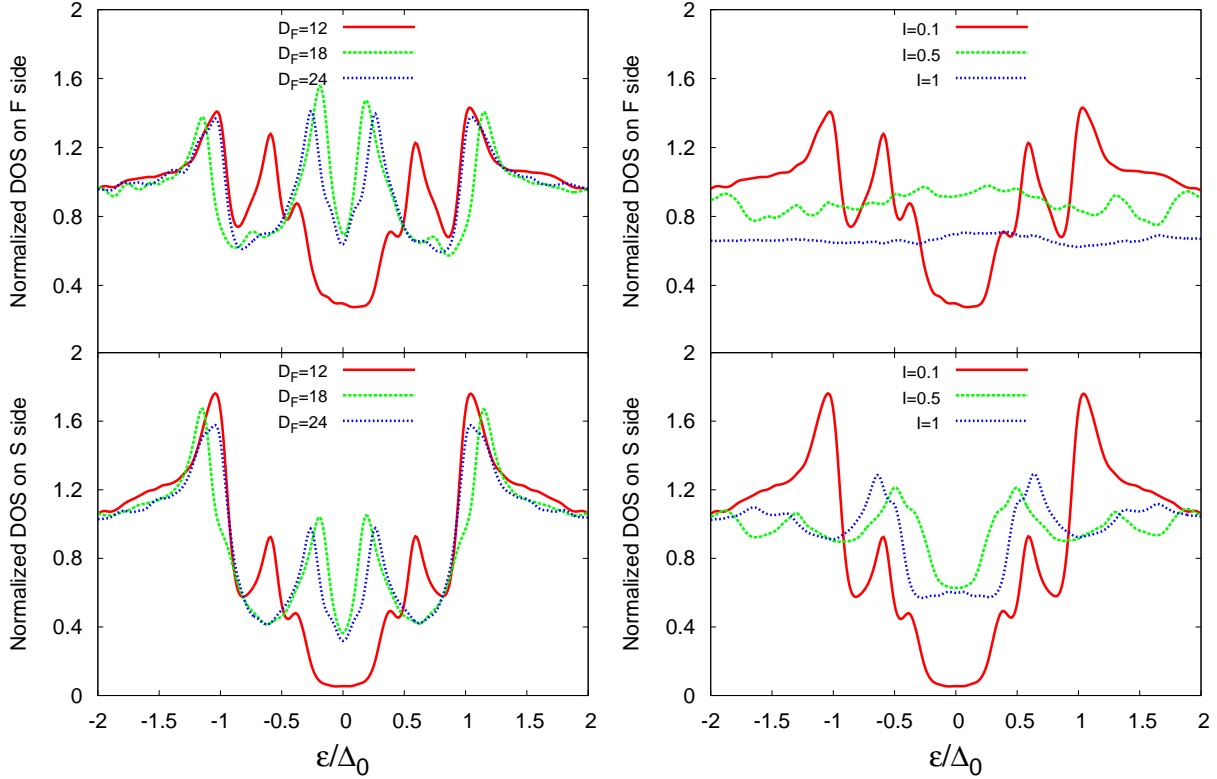}
\caption{(Color online) LDOS averaged over the $F$ regions (top panels)
and $S$ regions (bottom panels), plotted vs. energy. On the
left panels, the integrated LDOS is shown for different $D_F$ and $I=0.1$. 
On the right panels, the integrated LDOS
is shown for different $I$ and $D_F=\Lambda$. In all cases, the
superconductor width is set to $D_S = 150$.
}
\label{fig12}
\end{figure*}

Next, we discuss other important physical quantities 
that are related to the proximity effects including 
the local magnetization, ${\bm m}(y)$, and the 
the local DOS (LDOS). 
Considering that the ferromagnetism can drastically 
alter the superconductivity, one might wonder  about
the opposite case: how the local magnetizations 
behave near the $F/S$ interface.  
These so-called reverse proximity effects have
been studied 
\cite{hv3,cite:khov1,fryd,koshina,bergeretve,us69,cvk1} 
for a number of multilayer $F/S$ configurations 
with uniform exchange fields in each magnetic layer. 
Here,
the space-varying exchange fields in $F$ oscillate in 
the $x-z$ plane and 
are constant along the $y$ direction (see Fig.~\ref{fig1}). 
We computed the local magnetizations using 
Eq.~(\ref{mm})
for $D_S=150$, $D_F=\Lambda$ and three different values of $I$. 
The results are normalized in the usual\cite{us69,cvk1} way
so that for a putative bulk $F$ material with a
uniform internal field  characterized by 
the parameter $I$ the quantity plotted would
have the value $[(1+I)^{3/2}-(1-I)^{3/2}]/[(1+I)^{3/2}+(1-I)^{3/2}]$.
In 
Fig.~\ref{fig11},
each component of ${\bm m}$
is shown in a separate panel and their behavior plotted 
throughout the whole spiral magnet region and
some distance into S near the interface. 
Consider first the $x$ component: 
The corresponding component of the internal field (see 
Eq.~\ref{exchange}) vanishes at the outer interface ($Y=0$)
and goes smoothly to
zero at the $F/S$ interface which in this case 
is at $Y=D_F=\Lambda$. As a consequence, 
one can see that the $m_x$ component
undergoes a full period of oscillation in the $F$ material. 
The maximum and minimum values as a function of $I$ are
numerically what they should be, given our normalization.
However, as 
Fig.~\ref{fig11} clearly shows,  
the self-consistently determined $m_x$ 
does not vanish at the $F/S$ interface
and instead penetrates a short distance inside $S$.
This is a manifestation of the reverse proximity effect.
For the other transverse component, $m_z$ the situation is 
more complicated. The field component $h_z$, out of phase with
$h_x$, does not vanish smoothly at $Y=0$ nor 
at $Y=D_F$. Therefore the corresponding $m_z$ component
in $F$ is squeezed, and in addition to the  peak
at $Y=\Lambda/2$, which has the  expected location and value,
there are two smaller peaks at intermediate values. At the 
interface between materials, penetration of this component 
is appreciably more considerable than for $m_x$. 
The longitudinal component $m_y$, which is
induced by the uniform $h_y$ component, behaves qualitatively
as transverse components do
in  uniform ferromagnet $F/S$ structures.\cite{hv2}
Penetration into the $S$ layer occurs over a relatively
short distance, except at the smallest value of $I$ where
it is relatively larger although the overall scale is of course
smaller. The value of $m_y$ in the $F$ layer is again the 
expected one, consistent with our normalization. 

Finally,  we wish to
discuss  the LDOS. Here we will present results for 
the LDOS,  as
given in Eq.~(\ref{dos}), summed over spins, integrated over either the $F$
or the $S$ layer, and normalized, as usual, to its value
in the normal state of
bulk $S$ material. The results are given in Fig.~\ref{fig12}
where the energy scale of the horizontal axis is in
units of 
the superconducting gap
of bulk $S$ material, $\Delta_0$. 
The left panels of Fig.~\ref{fig12} show the  LDOS integrated over 
the $F$ (top) and $S$ regions  (bottom) for $D_F=\Lambda$, $1.5\Lambda$, 
and $2\Lambda$.
The superconductor  has a thickness $D_S=1.5\Xi_0$ and 
$F$ has a relatively weak 
exchange field, $I=0.1$.
For $D_F=\Lambda$, one can clearly see, for the 
integrated DOS in the $S$
side, peaks near $\varepsilon/\Delta_0=\pm 1$  as in the ordinary bulk
spectrum. There is additional subgap structure including 
proximity induced bound states at smaller energies followed by 
a very deep dip-nearly 
a minigap. Overall, the DOS structure
contains traces 
of the familiar DOS for a pure bulk superconductor.  
On the $F$ side, the integrated LDOS at this value
of $D_F$ still exhibits BCS-like peaks at $\varepsilon/\Delta_0=\pm 1$ 
and subgap dip, but the whole structure is much weaker
and the depth of the dip much smaller.
It is indicative 
of the superconducting 
correlations present 
in the
$F$ region.
In contrast, the subgap superconducting features in the integrated LDOS 
for larger $D_F$ values ($D_F=1.5\Lambda$ and $2\Lambda$)
are much less prominent, although the
peaks near $\varepsilon/\Delta_0=\pm 1$ remain. Nonetheless, there
are  still  shallow 
and discernible signatures 
in the gap region,  on both the $F$ and $S$ layers. 
For these two larger values of $D_F$ the results (as
compared on the same side) are
remarkably similar. This is surprising at first since
we have already seen that 
$T_c(D_F)$ has in this range of $I$ maxima  
near $D_F=\Lambda$ and $D_F=2\Lambda$ and a local minimum 
at $D_F=1.5\Lambda$ as shown in Fig.~\ref{fig3}.
From that, one might naively guess 
that the integrated LDOS for $D_F=2\Lambda$ should behave as that
at $D_F=\Lambda$,  with a different integrated LDOS for 
$D_F=1.5\Lambda$. This expectation
is incorrect because, as one can see on a closer inspection of
Fig.~\ref{fig3},  $T_c$ at $D_F=\Lambda$ is
higher than that at $D_F=2\Lambda$ although both are near local
maxima. Furthermore,  $T_c$ at $D_F=1.5\Lambda$ is closer
to the $T_c$ value at $D_F=2\Lambda$ than to that at $D_F=\Lambda$. 
Since  $T_c$ is associated with
the magnitude of the singlet pair amplitudes, in which the LDOS is indirectly 
correlated to, one should conclude that the LDOS corresponding to $D_F=2\Lambda$ 
should be similar to $D_F=1.5\Lambda$ rather than $D_F=\Lambda$.
Indeed, the results confirm this notion. 

On the right panels of Fig.~\ref{fig12}, 
we present the integrated LDOS on both
the $F$ and the $S$ sides for different exchange fields, $I=0.1$,
$I=0.5$, and $I=1.0$, at $D_F=12=\Lambda$. We see that when
$I$ is increased from $I=0.1$,
the integrated LDOS on the $F$ side becomes
quite flat (at the value 
$(1/2)[(1+I)^{1/2}+(1-I)^{1/2}]$ as
per our normalization) and essentially devoid
of a superconducting signature. On the 
$S$ side, the integrated LDOS at $I=0.5$
and $I=1.0$  still retains some vestiges of
the structure seen in the $I=0.1$ case.
However, the integrated LDOS at $I=0.5$ 
on the $S$ side is slightly different than that at $I=1.0$.
The dip in at $I=1.0$  is wider
than for $I=0.5$, in a way 
more superconducting-like. What happens is that at larger values 
of $I$ the mismatch between the Fermi wavevector in $S$ and the
Fermi wavevectors in the up and down bands in $F$ increases.
This diminishes the penetration of the Cooper pairs
into $S$ and hence the overall scale of the proximity
effects. We recall that the overall dimensionless scale
of the proximity effect in $F$ is roughly $\Xi_F=1/I$. 
Consequently, superconductivity is impaired in $S$
over a smaller scale when it is in contact with 
a stronger ferromagnet.
Having said that, one might argue that at $I=0.1$ the integrated LDOS 
on the $S$ side should have a smaller dip than the other two curves
for stronger $I$. However, we have to consider here also the
overall behavior of the $T_c$ vs. $I$ curves at constant $D_F$. This
behavior is once again oscillatory but with a superimposed decay.
The overall decay results in $T_c$ being higher at $I=0.1$ than at either
$I=0.5$ or $I=1$, but the oscillations produce a higher value of $T_c$
at $I=1$ than at $I=0.5$. This explains the progression of the curves.
All the above discussion and results indicate that the LDOS can 
provide, if properly analyzed, another perspective and 
additional information on  
the superconducting nature of our bilayers. 

\section{Conclusions}

We have studied several aspects of proximity effects
in $F/S$ bilayers, where the ferromagnet has a 
spiral structure characteristic of
rare earths such as Ho, by numerically solving the 
self consistent BdG equations.
We have calculated $T_c(D_F)$, the critical temperature as
a function of magnet thickness, for different 
parameter values.
The  $T_c(D_F)$ curves exhibit a
fairly intricate oscillatory behavior
which 
is found to be related to both the strength $I$ (as
they would for a uniform magnet) and 
the periodicity $\Lambda$ of the spiral exchange fields inherent in 
the magnet.
As is the case for $F/S$ structures in which $F$ 
is uniform, we observe  reentrant 
behavior with $D_F$ when $I$ is strong enough. 
The physical reason behind this $D_F$ reentrance in our bilayers 
is similar to that in  ordinary $F/S$ structures but the additional 
periodicity associated with the magnet, which in many
cases dominates the oscillations,  makes the behavior more
complicated.
As a function of $D_S$, we find that $T_c(D_F)$  
can also exhibit  $D_F$ reentrance even at small $I$ when
$D_S$ is of the order of the superconducting coherence length.
The additional oscillations produced by the magnetic
structure lead also to pure reentrance with temperature:
superconductivity occurs in a finite temperature range
$T_{c1}<T<T_{c2}$. An example of this
reentrance at a very
small $D_F$  ($D_F\sim 0.5\Lambda$) was previously\cite{cite:cvk} presented.
Here we report that this reentrance
can also occur when $D_F>\Lambda$, where it
should be experimentally easier
to realize. To elucidate the physics
underlying these reentrant phenomena, we
have evaluated the singlet pair amplitudes and thermodynamic functions.
The competition between condensation energy and entropy
is responsible for the $T$ reentrance: the
superconducting state may be, under certain circumstances, 
the high entropy state, leading to
recovery of the normal state as $T$
is lowered. The calculated
thermodynamic quantities are fully consistent with 
the $T_c(D_F)$ phase diagrams
and the singlet pair amplitudes.

When the magnet
has a spiral structure both 
the $m=0$ and $m=\pm1$ odd triplet components can be induced simultaneously.
This is not the case in uniform-magnet bilayers:
at least two uniform  misaligned $F$ layers are needed to generate 
the $m=\pm1$ component. 
We studied the odd triplet pair amplitudes in our bilayers,
and found them to be long-ranged in both 
the $S$ and $F$  layers. 
We have analyzed the time delay dependence of the odd triplet
amplitudes. The results are consistent
with our previous work on both $F_1/S/F_2$ and $F_1/F_2/S$ trilayers,
but the additional $\Lambda$ periodicity leads to important
differences. We characterized the triplet long range 
behavior by introducing the appropriately defined lengths. 
We found that the relevant
proximity length oscillates with $D_F$ and these oscillations
depend on the strength and
periodicity of the exchange field. Our methods are
likely appropriate for many experimental
conditions,
as evidenced 
by the  consistency of our results 
with recent tunneling experiments.\cite{robinson} 

We have also considered the reverse proximity effects: the influence of the
superconductivity on the magnetism.
We found all three components of the local magnetization penetrate
in slightly different ways 
into the $S$ layer. At larger $I$ this is a short-ranged phenomenon,
but it is otherwise for weak magnetism.
Both $m_x$ and $m_z$ oscillate in the $F$ regions to reflect
the spiral exchange field. Finally, the calculated LDOS reveals important
information and discernible signatures 
linked to the proximity effects in these bilayers and are
correlated to the superconducting transition temperatures.

\acknowledgments

We thank C. Grasse for technical support. K.H. is supported in part by
ONR and grants of computing resources from DoD (HPCMP). O.T.V. and
C.T. Wu are 
supported in part by IARPA under grant N66001-12-1-2023.

\end{document}